# Chapter 8

# Interconnected virtual space and theater: a research-creation project on theatrical performance space in the network era

## 1.1. Introduction

Since 2014, we have been conducting experiments based on a multidisciplinary collaboration between specialists in theatrical staging and researchers in virtual reality, digital art, and video games. This team focused its work on the similarities and differences that exist between real physical actors (actor-performers) and virtual digital actors (avatars).

From this multidisciplinary approach, experimental research-creation projects have emerged, which we have previously described [GAG 15] [GAG 17]. These projects rely on a physical actor playing with the image of an avatar, controlled by another physical actor via the intermediary of a low-cost motion-capture system (a mobile inertial motion-capture system, which we will describe later).

In the first part of this paper, we will introduce the scenographic design on which our presentation is based, and the modifications we have made in relation to our previous work. Next, in the second section, we will discuss in detail the impact of augmenting the player's game using an avatar, compared to the scenic limitations of the theatrical stage. In part three of the paper, we will discuss the software-related aspects of the project, focusing on exchanges between the different components of our design and describing the algorithms enabling us to utilize the real-time movement of a player via various capture devices. To conclude, we will examine in detail how our experimental system linking physical actors and avatars profoundly alters the nature of collaboration between directors, actors, and digital artists in terms of actor/avatar direction.

## 1.2. A multidisciplinary experiment involving live performance and digital art

The introduction of an avatar controlled in real time by an actor on a theater stage mobilizes the transdisciplinary skills of the team in the fields of theater, digital art, and video games. It leads to the definition of new terms and functions having to do with the expressive potentialities explored. The experimental system is continuously evolving in terms of the improvement of the software and motion-capture devices used, eventually resulting in precise documentation and a permanent adjustment of the concepts and technical solutions necessary for truly shared experimentation.

### 1.2.1. *Definitions of avatar and mocaptor*

First of all, we must define some concepts. While the idea of an avatar itself has been adequately defined [DAM 97] [SCH 12], it remains to define the concept of a mocaptor, a neologism derived from fusing the terms "motion-capture" and "actor". A mocaptor is a physical actor equipped with a motion-capture device, which enables the real-time control of an avatar generated as a synthetic image. While these techniques, frequently used in Hollywood [BAL 09], rely on large teams of computer graphics artists in charge of motion processing, the approach we use, due to its real-time nature, is closer to the work of Patoli [PAT 10]. Moreover, our approach is based on a real-time dynamic with little or no cleanup work; that is, without correcting in post-production measurement errors or noise generated by the capture process. At most, we use basic filters such as the Kalman filter to eliminate major capture errors.

### 1.2.2. *Description of system*

For our experiments, we have developed an improvisation system for theater students to use during rehearsal. This system has been successfully used for teaching purposes since the 2016/2017 academic year.

The stage space defined in the project is occupied by several entities (see Figure 1):

A) Physical actors, located in the center of the stage;

B) The mocaptor controlling an avatar;

C) The digital artist(s) and participant(s) involved within parameters to be defined;

D) The avatar(s) represented onscreen.

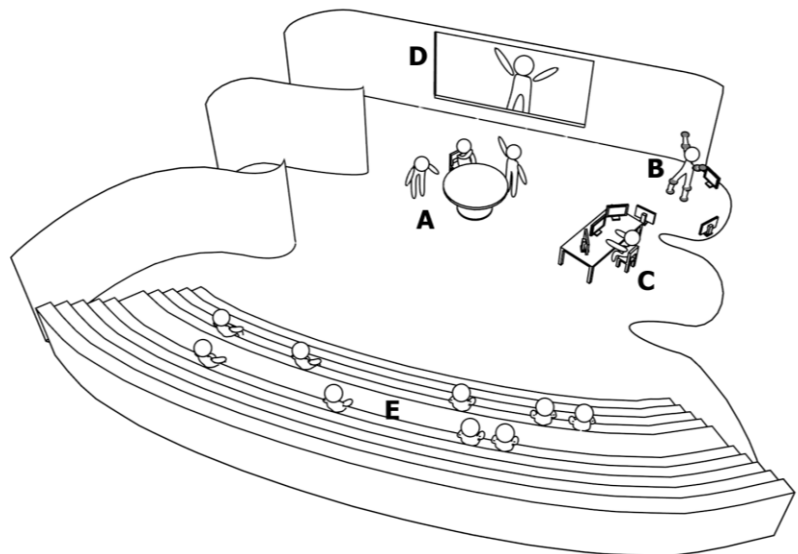

*Figure 1. Diagram of staging in practical workshops*

From a performance perspective, a space (E) is reserved for the audience, where the director is usually located as well. We might think at first that scenographic actions concern only the relationship between the physical actors (A) and the avatar (D); however, the avatar is the combined result of the actions of the mocaptor (B), the artist, and the digital participants (C). Thus we have chosen to put the three entities B, C, and D in the stage space, leaving it to the viewer to choose to concentrate his gaze on whatever interests him the most. This offers the actors and the director the ability to increase the number of staged interactions among all participants.

### 1.2.3. From Kinect to Perception Neuron: a new mobility

One development that has occurred since [GAG 17] is the shift from Kinect to a combination of inertial motion capture called Perception Neuron Mocap. This system is composed of a network of 32 inertial measurement units (IMUs) connected in a hierarchy and communicating with the computer via WiFi connection. This system allows us to capture the accelerations and angular velocities of certain parts of an actor's body (head, shoulders, elbows, hands, fingers, torso, knees, and feet), which we then use to recreate the motion captured. This mechanism was originally designed for low-cost, real-time motion capture for use in video games.

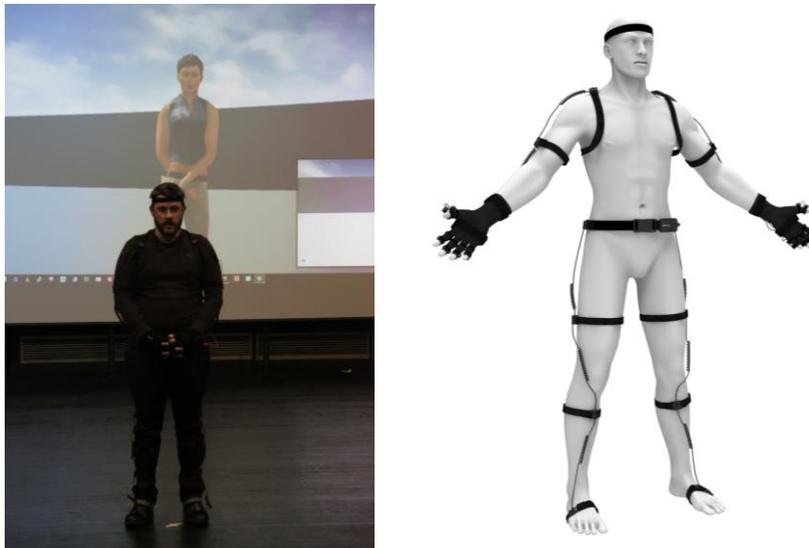

*Figure 2. A mocaptor and the avatar reproducing his gestures (left) and the capture device (from the neuronmocap.com, manufacturer's image) (right)*

With Perception Neuron, 1) the area of play becomes much larger, since the mocaptor is no longer limited to the Kinect's capture zone, but rather to the capture zone of the WiFi terminal (cf. Figure 3); and 2) the mocaptor has greater freedom of movement, as he/she is no longer required to act facing the Kinect. Finally, the quality of the motion-capture is much higher with Perception Neuron. However, this system must be used with great care so that it does not lose the position of the body, as it functions relative to points of contact on the ground.

The possible juxtaposition of the mocaptor's performance space to that of the physical actors allows for a stimulating triangulation between physical space and digital space. An actor can actually have a double presence; a digital one on the screen and a physical one on the stage. However, this triangulation assumes the specific organization of a double interaction: that of the mocaptor directly with the physical actors, and that of the mocaptor with the physical actors via the intermediary of the avatar. Faced with this complexity, we chose to begin our experimentations by confining the performance space of the mocaptor to a delineated area outside the stage space of the physical actors (see the upper right corner of Figure 1). However, compared to our previous experimentations with Kinect, the mocaptor equipped with Perception Neuron can cover a larger space by turning back on itself. On the other hand, the question of returns is still significant, which explains the positioning of three screens around the mocaptor.

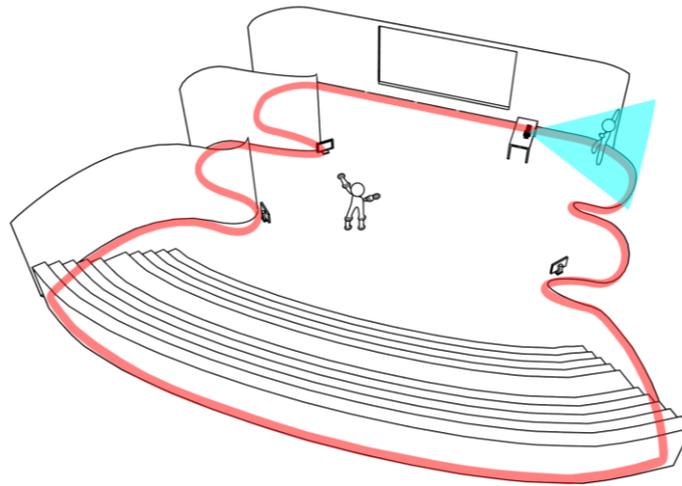

*Figure 3. Action zones of Kinect (blue) and of Perception Neuron (red)*

### 1.3. Acting relationship between the mocaptor and the avatar

The end purpose of the avatar is to construct a acting situation with a physical comedian. To do this, it must occupy a digital space that is linked to the physical performance space of the actor. The reciprocal positions of the two entities in their respective spaces impose a field of limitation within which to build what we call the spatial disposition of a character (physical or digital); that is, the place toward which it projects a scenic (in-performance) action.

#### *1.3.1. Controlling the avatar's spatial disposition*

The first challenge consists of positioning the avatar evolving in a 3D space, but present in the physical space via a two-dimensional image projected on a video projection screen facing the audience behind the physical actors (cf. Figure 4 - N°1).

If the viewer is located in the center of the audience, where the director usually sits (space E), he has a completely different perception from that of the mocaptor located in the lateral space (space B) due to the effects of perspective. The mocaptor cannot thus base his movements on the feedback he sees on the video projection screen to orient his avatar in relation to the actor. He must follow the director's instructions and must incorporate a "bias" against his own perception into his actions. We have already noted this limitation in our previous experiences with Kinect [GAG 17], and we have solved it for now by finding a fixed orientation for the mocaptor facing his Kinect and the video projection screen at the same time, which enables him to have a lateral perception of the actor's physical performance space, and to learn the biases "necessary" for establishing the avatar's contact with its lateral partner according to the directions of the director positioned among the audience members.

The constraint of being positioned facing the Kinect (see below) has made it possible to set limits on the "biases" to be internalized that are acceptable for the mocaptor, who is now able to ensure that his own performance space corresponds with the digital space of the avatar in relation to the actor acting in front of the screen.

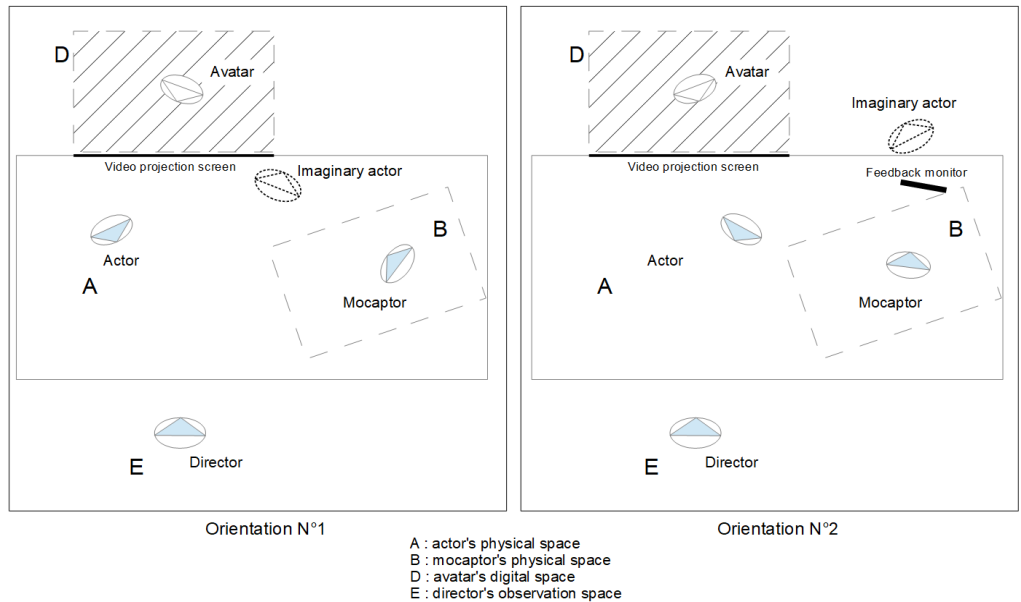

*Figure 4. Controlling the avatar's spatial disposition via mocaptor movement*

### 1.3.2. The mocaptor's reference space: the Ninja Theory example

We must address the complex question of articulation among three performance spaces in order to better understand the limitations on the mocaptor's acting. This performs in his physical performance space (B), which allows him to exist in a digital performance space by controlling the figure of the avatar (D), which must position itself in its turn in relation to the acting of the actor evolving in a third space (A). To better understand the specific characteristics of our system, we will compare it to the pioneering work of Ninja Theory and Epic Game on the cutting edge of video games and computer-generated 3D film [SIG 16]. The demonstration offered by Ninja Theory consists of constructing a trailer for the video game Hellblade in real time and in a condensed production pipeline.

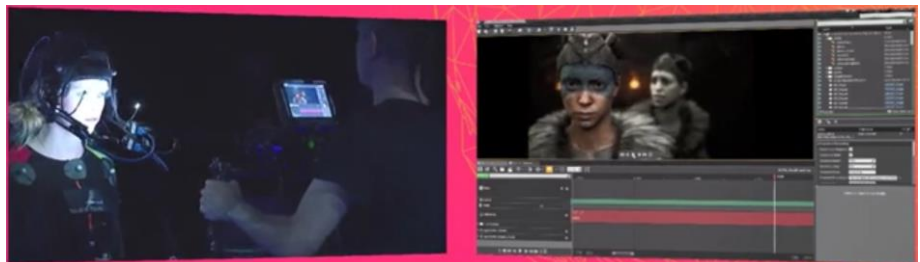

*Figure 5. The real-time capture device used by Ninja Theory and Epic Game*

A female mocaptor is equipped with a double system of facial and body motion capture which enables a realistic real-time rendering thanks to specially designed motion retargeting rigs. In front of a mirror, the avatar Senua speaks to her double, located in another world. Senua and her double are two independent characters. A cameraman (Figure 5, left) uses a rendering camera and a rig to determine in real time the camera position that creates the scene in the video game engine (Figure 5, right). The first take consists of filming the mocaptor's first performance; she plays the role of the specular double, who speaks into the mirror to the true Senua. Then the double walks forward to pass through the mirror and continue talking to Senua while moving in a half-circle around her. The performance is done by the mocaptor using pre-set marks on the floor and internalizing the overall temporalities of the dialogue. The take is recorded in the game engine and can be replayed with various effects used as desired on the character and the performance space. Next, a second take is shot to incorporate the character of Senua (see Figure 5, right, which shows the juxtaposition of the two characters, though only one is captured). The cameraman specifies key positions to the mocaptor, particularly at the point when her double passes through the mirror and at the moment when she goes through the mirror herself to enter her double's world. The previous sequence is begun with the character of the double, which has just been recorded, and the mocaptor acts the role of the true Senua in real time and reacts to the double's actions, saying her parts of the dialogue in the spaces of silence left during the first take. The cameraman adjusts his camera angle in real time according to the relationship between the two characters, and this is where his artistic and technical prowess comes into play. The cameraman can actually compose his shot as if he were filming two real film actors, even though they are "only" virtual avatars.

This experiment demonstrates a promising view of future real-time pre-visualization. In this context, the avatar's digital space mingles directly with the mocaptor's performance space. The mocaptor projects herself emotionally into a character with whose realistic expressive potentialities she is familiar. For the first take, she imagines that she is facing the other character, whom she will play during the second take and whose marks on the floor she can make note of. Then, during the second take, her scene partner, though invisible to her, is visible to the audience and the cameraman. The third performance space of the physical partner also blends with the two other ones. It is the responsibility of the cameraman-filmmaker to give positioning directions to adjust control of the avatar according to the pre-recorded game path. The precise superimposition of the three performance spaces thus facilitates the mocaptor's acting.

### 1.3.3. A closer look at the articulation of the reference spaces

In our theatrical trial, the three spaces are sharply separate, which causes difficulty for the mocaptor's acting. The digital performance space of the avatar (D) is represented via a flat image giving the impression of another physical space contiguous to that of the actor (A) (cf. Figure 4). In order to interact correctly with his physical partner, located in front of the video-projected image of the avatar's space (video projection screen), the mocaptor must mentally transpose the partner's movements in relation to the digital space of the avatar in his own performance space (B). If we assume that the mocaptor is looking at the video projection screen and orients himself correctly with his avatar, with the help of the director located in space E, in order for the avatar to interact with the actor, it is the responsibility of the mocaptor to imagine that he is looking at an imaginary actor located at the same distance from him as the real actor in relation to the avatar. Thus, if the actor moves in his space and takes up a new position (cf. Figure 4 - Orientation N°2), the mocaptor needs to imagine the same movement by the imaginary actor and position himself accordingly so that his avatar will also be correctly positioned. This results in his no longer being able to see the video projection screen, and a feedback monitor must be installed so that he can continue to have contact with the image of his avatar.

In practice, this imaginary transposition of the real actor's movements is, of course, impossible for the mocaptor to accomplish; he would have to do geometric calculations in his head about movements taking place more or less behind him. The practical solution used in our platform consists of turning once again to the director, who indicates the correct movements to the mocaptor, who can memorize key positions, which he connects to the positions taken by the avatar, with which he maintains visual contact via the feedback monitor, and he empirically constructs his own course of movement as an echo of the actor's movements, which he observes on the physical stage A.

### 1.3.4. Mobility of the mocaptor in his performance space

In order to make the mocaptor's work simpler, we have developed a second method inspired by video game motion retargeting techniques. Instead of asking the mocaptor to move, we play on the position in space B of the mocaptor's acting, by applying the necessary rotation to it so that the avatar is correctly positioned in relation to the actor in space A (cf. Figure 6).

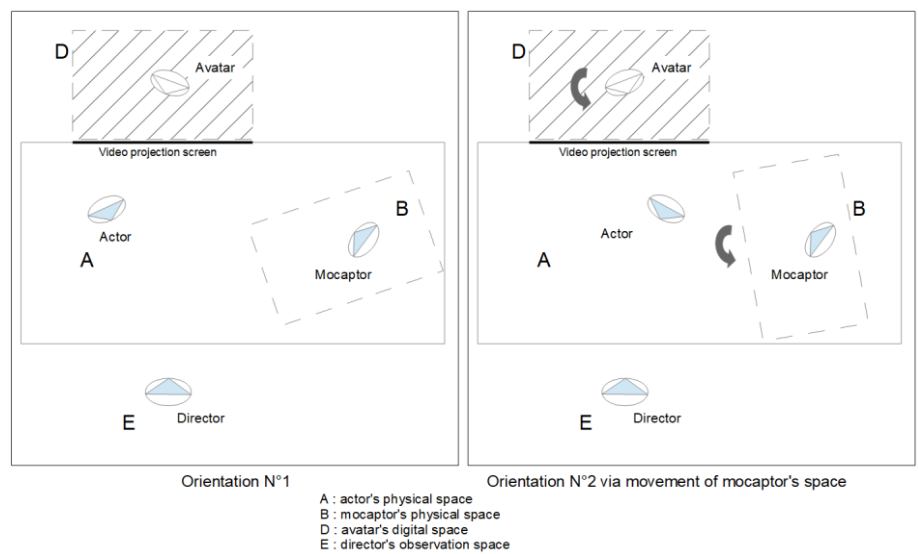

*Figure 6. Controlling the avatar's spatial disposition via rotation of space B*

In reality, this rotation of space B is only an image manifesting an algorithmic operation we carry out on the avatar in space D, and which we will discuss in detail in section 1.4. One clear advantage is immediately noticeable: the manipulation enables the mocaptor to maintain contact with the video projection screen. However, as it turns out, it is necessary for the mocaptor to retain freedom of movement, and orient his body in positions that cause him to lose visual contact with the video projection screen. Thus we have placed monitors in his performance space B that augment the yield of the video projection screen and allow the mocaptor to maintain a relationship with the final avatar (cf. Figure 1 – space B). This need for feedback is similar to an opera singer's need always to be able to see the orchestra conductor. This is why, in an operatic production, several monitors are arranged around the performance space and the audiene space, in order to give the singer the freedom of movement required to follow the director's instructions. The singer can thus perform expressively without losing the lifeline to the orchestra conductor, which guarantees musical cohesiveness between the singer and the orchestra.

The introduction of digital manipulations to the mocaptor's motion-capture information will have significant consequences for the creative potentialities of the system and the relationships among the various artistic collaborators, several aspects of which we will analyze in the last part of this chapter.

## 1.4. From mocaptor to avatar from a technical perspective

In this section, we will give a more detailed description of the control conditions of the avatar (D) by the mocaptor (B) in the context of a staged interaction of this avatar with physical actors (cf. section 1.3). In reality, there are three major issues to be dealt with:

- The transfer of a real actor's movement to a virtual character while taking into account differences in size and morphology that may be encountered in mocaptors and virtual avatars;

- The combination of multiple control sources (mocaptor and digital technician) to manipulate and change the avatar;

- The necessity of doing all of this in real time in an adaptable evolutive architecture.

### 1.4.1. Two-stage motion retargeting

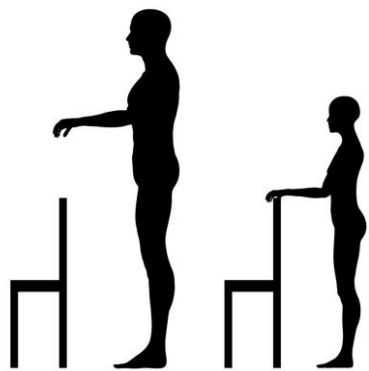

*Figure 7. Adapting an action to the actor's morphology*

To understand the necessity of effective motion retargeting, we can see that a difference in morphology, or body type, between two people causes numerous differences in gestures (though these may be technically identical) (cf. Figure 7). Because we may have mocaptors and avatars with different body types, we have developed a two-stage processing procedure that allows us to transfer a mocaptor's movement on to a complex avatar, with the help of a third party: the puppeter.

The concept of puppeter is closely related to the original definition of the term puppeteer, but goes well beyond this definition, as we will explain in detail below.

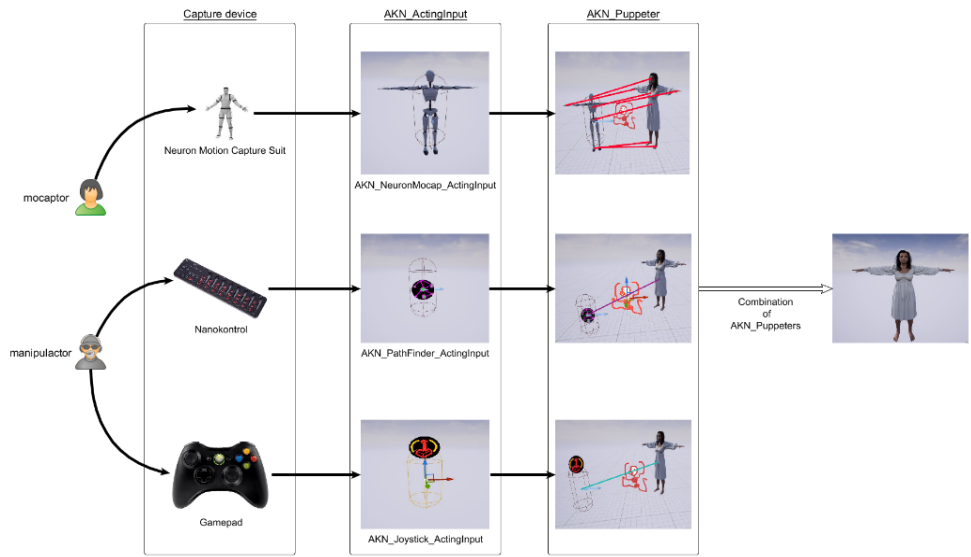

*Figure 8. The 2 phases of motion retargeting with 3 capture devices*

1. The mocaptor manipulates the motion-capture device (Perception Neuron, Kinect, etc.), which sends data via WiFi to the computer, and the data sent by the motion capture device is applied to a neutral character. This allows us to have formalized values; the neutral character is not visible to viewers.

2. The puppeter executes a second motion retargeting on complex avatars that often do not have a much more evolved architecture in terms of positioning or number of joints.

It is important to understand that the puppeter is not a marionettist/puppeteer manipulating the different parts of the avatar's body, but rather a "blender" of motion sources for these different parts, which the puppeter can then combine and rearrange according to the motion retargeting issue, but also dynamically or even in terms of behavior. This area has already been explored in an on-set pre-visualization tool intended for film, called the OutilNum project [PLE 15] and using the concept of "behavior reproduction", which will eventually be incorporated into the new architecture.

The act of executing motion retargeting in two stages has several advantages. First, the neutral character acts as an initial filter able to eliminate excess data, delete unnecessary data, and make data uniform, no matter what motion capture device is used. In addition, using a neutral character with well-established morphology allows us to retain and reuse animations for future use. Profound differences may exist in the architectures of a complex character created on an ad hoc basis. Because the architecture of the character's joints is not homogeneous, it is impossible to transfer the movements of one complex character to another character. By animating for the neutral avatar using the puppeter, we can make the most possible use of animations. Finally, the fact that the source movements are all applied to the same character enables us to combine them with each other in order to obtain unique animations.

### *1.4.2. Avatar movement: combination of multiple sources*

The operation of positioning within the mocaptor's reference space mentioned at the end of section 1.3 is carried out as follows. The NeuronMocap data is collected in the AKN_NeuronMocap_ActingInput tool, which is used to animate the neutral character. The AKN_NeuronMocap_ActingInput tool then serves to control the complex avatar using the Puppeter object.

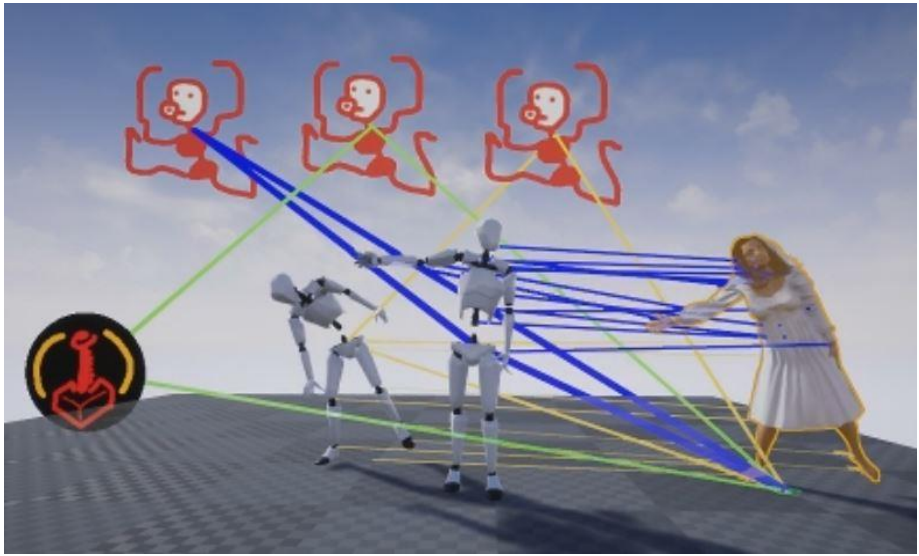

*Figure 9. NeuronMocap_actingInput, joystick_input, and puppeter in action*

However, we can also use other information to control the avatar, generated by a gamepad, for example. We then use the AKN_RefMove_ActingInput tool to recover the information from the gamepad device and use it to modify the mocaptor's space B, which goes back to combining this new data with the data from the AKN_NeuronMocap_ActingInput tool using a second puppeter working on the complex avatar. Figure 9 particularly shows the possibility offered by the architecture of combining NeuronMocap_ActingInput with two different mocaptors to control the final complex avatar. Note as well the presence of a Joystick_Input tool, which can be used to change the avatar's position with a gamepad. While the NeuronMocap_ActingInput data is produced by what we have called the mocaptor, we have chosen here to refer to the digital participant who uses the gamepad to position the avatar as a "manipulactor".

The devices used by a manipulactor can include the gamepad, which is used to control the position and movement of the final avatar, as in a video game; as well as a midi controller such as the Korg NanoKontrol2, which is useful in triggering effects using potentiometers, cursors, and on-off buttons (cf. Figure 8). Keyboards and mice may also be used, as well as any other controllers that allow an ergonomic relationship between the human body and the machine producing digital parameters.

### 1.4.3. Combination with independent behavior: pathfinding

The manipulactor may choose to use pathfinding functionalities to facilitate the mocaptor's performance in the avatar's digital space. Our pathfinding algorithm is based on classic shortest-route algorithms that have been used for many years in video games. A navigation mesh (navmesh) is generated according to the topology of the stage and the obstacles placed on it, and we use an A* algorithm to traverse the graph, using the distance between the character's position and the navigation points attached to the navmesh as a heuristic function [CUI 11]. The AKN_Pathfinder_ActingInput tool, combined with a puppeter, is used to guide the complex avatar independently in the digital space D, helping it to avoid any possible obstacles. Our practical experiments have in fact led to the use of a digital game space for the avatar that is much more capacious than the mocaptor's performative space. The pathfinding functionality enables us to organize intelligent movement for the avatar, independent of the mocaptor's movements.

The taking-over control from the mocaptor is decided upon by the manipulactor, which initiates the independent movement of the complex avatar. Then, control is released and the mocaptor once again has total control of the avatar. At the moment of release, the manipulactor must be careful to ensure that the complex avatar is correctly positioned in the new performance location, and possibly use a gamepad to reposition the complex avatar. It should be specified that the mocaptor retains control of the avatar's limbs during the pathfinding phases, which involves good coordination with the manipulactor to choose the timing of control-taking and disengagement.

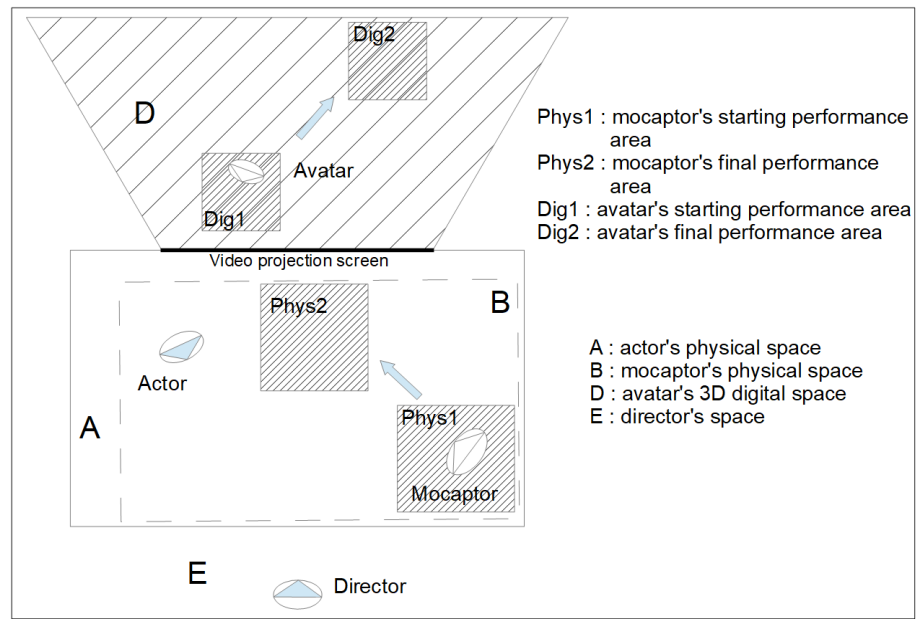

*Figure 10. Scenario in which pathfinding is used to guide an avatar*

A second use of pathfinding has proven necessary in situations where the mocaptor performs in the physical space A of the actor. The stage direction concept was to duplicate on the video projection screen, in a specific digital space, the actions of the mocaptor-actor performing in the physical performance space representing another space. Figure 11 represents the superimposition of the actor's space A with the mocaptor's space B. The mocaptor thus plays a double role; he is simultaneously acting and controlling an avatar. The avatar's position is represented in the cross-hatched virtual digital space D. Space D is shown approximately in perspective according to the point of view of the director, located in the viewer observation space E. For reasons of staging, there is no 1:1 scale homothety between the two performance spaces B and D; therefore it is not possible to make the mocaptor's performance space correspond to that of the avatar. It has thus been decided to choose specific action zones in each of the two spaces of a few square meters each in size, and to move from one zone to the other using the pathfinding function in the digital space. The mocaptor moves from zone Phys1 to Phys2 in his physical space B, while the pathfinder transports the avatar from zone Dig1 to Dig2 in the digital space, under the control of the manipulator. Note that the manipulactor must reorient the avatar's final position based on that of the mocaptor so that the avatar maintains the spatial disposition desired by the director. This means that it is necessary to construct a final location for the avatar, which is implemented by the manipulactor at the moment when the pathfinder releases its control of the avatar, using a preset system utilizing the AKN_Event tool (cf. 1.4.4), which allows the coordination of avatar control information circulation.

We would like to specify our reasons for choosing a combination of the terms 'manipulator' and 'actor' to describe the function of the mocaptor's partner, who extends the mocaptor's movements in the final complex avatar. We believe that the manipulactor must draw on the theatrical expressiveness of an actor's performance to successfully embellish the mocaptor's movements. It is a matter of initiating transformations in physical appearance that will be interpreted by a partner actor and the audience. Like the actor-marionnettist pulls the strings of his marionettes in keeping with the art of dramatic acting, the manipulactor must embellish the mocaptor's performance sufficiently. We will look at the consequences of the emergence of these two roles in section 1.5.

### 1.4.4. An architecture oriented toward interconnected objects

The current software architecture is descended from a development library begun in 2002, called Taupistool [PLE 07]. It was actually necessary for us to carry out simultaneous studies of artificial intelligence for visual effects in films, and interactions with virtual reality devices for sound detection and image analysis for artistic installations and effects systems for live performances and pedagogical support. This library, programmed in C++, was interconnected with multiple 3D platforms (Maya, Virtools, Motion Builder, Ogre, and a real-time 3D engine developed internally) and enabled the digital artist to reuse developments, either for real-time or precalculated movements. This architecture subsequently evolved and was redubbed AKeNe (dandelion pollen) to mark both its creation within the INREV EA4010 EDESTA Paris VIII team (via an homage to its founders Michel Bret, Edmond Couchot, and Marie-Hélène Tramus, who recreated the interactive work "Dandelions" in 2005 and its capacity to disseminate itself. This adaptable framework was then deployed in the OutilNum project mentioned above, as well as in a number of systems mixing theater and virtual reality [GAG 15]. Finally, it was restructured as a network so that it could be shared on multiple machines, creating different work stations.

It has also been our intention to improve communication among the various control stations at events (keyboard entry, midi controller, etc.). This relies on a system of simple discrete events (SDEs) which are non-centralized and have no wait queue [CAS 98], which we have called AKN_Event_Manager. We assume that there is little or no delay between the transmitter of the event and its recipient. Likewise, it is highly probable that we will have to rethink this mechanism if the network output among the various stations is too low. This is possible, as the information transmitted is very small in size (less than 10 ko) and, except for data generated by the combination of motion captures, is intermittent. The message transmitted relies on an archetypal boxing/unboxing system [LER 97] able to transmit simple data (float, string, etc.) as well as more complex data (objects and other types).

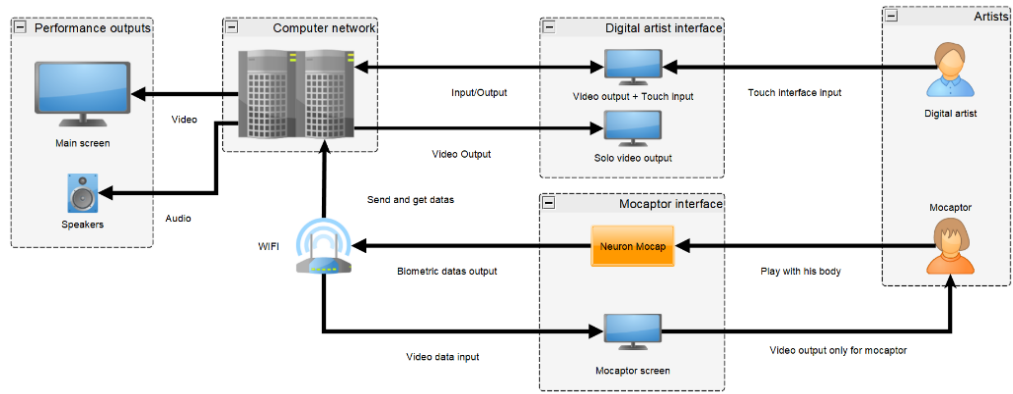

*Figure 11. Diagram of input and output according to local network and WiFi*

Though the mocaptor uses his body to send information via WiFi, the modifications made by the digital artist and the manipulactor are made directly on the computer (see Figure 12) by means of the keyboard, mouse, and various other devices (joystick, midi controller). Each of these work stations has one or more feedback screens to help with understanding actions taking place on the virtual stage D. The interface currently used by the digital artist has a number of screens showing various parameters that can be adjusted in real time. Thus, there are two categories of tools: those meant for use by all of the people present, and those meant to be used exclusively by the manipulactor(s) and mocaptor(s). The first category of tools is used to transmit aesthetic information to the computer and then to the main screen for the purposes of modifying the virtual representation. The second category of tools communicates past and future information in order to aid with directing. Though we have already put some basic tools in place, our intention is to create ergonomic communication systems between the mocaptor, manipulactor, director, and actors.

## 1.5. A practical application that raises new questions

Our experimentations have raised several questions from a theatrical, technical, and game design perspective. Firstly we will examine those having to do with the appearance of new protagonists on the stage; we will then address questions raised by the staging of virtual décor for the augmented stage, as well as the new role that could be played by digital artists in the theater.

### *1.5.1 New experimentation spaces for actors, directors, and digital artists*

The potentialities created by the taking into account of the human body's movements via a capture device transmitting via WiFi open up many stimulating creative perspectives for directors, but they also pose challenges in terms of ergonomics and team management. We have seen that the mocaptor projects rotations of parts of his body into the avatar, but the final control of the avatar is also in the hands of the manipulactor, who can influence certain parameters including the rotation and translation of the avatar, or its overall position, via pathfinding. The mocaptor can move in a given direction while his avatar goes in another direction determined by the manipulactor depending on constraints having to do with the physical stage space, notably the position of the actors with whom the avatar is interacting (cf. section 1.3).

The mobility gained by the mocaptor and the tools developed to insert his movements into a hybrid physical/digital scene through the intermediary of the manipulactor make it possible to create new combinations of movements in collaboration with the digital artist and the director. By acting to organize his performance space in relation to the digital space of the avatar, the mocaptor can thus plan to interact while playing two roles: that of the avatar's controller and that of the physical actor, in direct contact with the other physical actors in the system. This latter scenario would be of great interest for a more in-depth study in order to explore what the rules might be for a bodily connected user in relation to avatars which this user could manipulate in mixed reality to interact in other spaces.

Another creative route consists of enabling the mocaptor to control multiple avatars simultaneously. Each avatar makes the same movement as the mocaptor, but its spatial orientation is controllable. In this way we might recreate a Classical-era theatrical choir or a crowd of individuals with a single actor. As another example, the director might wish to have an avatar perform while aloft. The mocaptor could be asked to act as if he is taking flight, and then the manipulactor would take over by raising the avatar into the air, with the mocaptor then acting as if he is levitating.

However, the fact that the avatar's movement is the result of the combined control of the digital artist, the mocaptor, the manipulactor, and the director makes it necessary to create dialogue tools so that everyone can communicate. This leads more broadly to a redefinition of their collaboration in the context of an augmented scene, notably with regard to the challenges of "avatar direction", which corresponds to actor direction in the traditional theater.

### 1.5.2. The problem of visual composition for augmented scenes

The visual composition of an image is vital to ensuring its communicability [MAT 10], but also to allow aesthetic expression [KAT 91]. However, as we have seen in the previous sections, the visual composition of a virtual scene with avatars can change in many ways: the avatar turns and moves, in an environment in which the architecture can evolve, with a camera and lights which may also be in a continuous state of change. These different parameters raise the question of real-time virtual composition, for augmented scenes in particular.

The case of video games is already clear in this respect: visual composition is essential in these games for the ergonomic experience of the player, but must also guarantee pleasant reading of the image [SOL 12]. Thus it is common in game design to ensure the "three Cs" [ALB 15], for: Camera, Controls, Character. Real-time visual composition for video games is therefore planned to ensure that the player will have an optimal experience. Yet we can see a significant difference between composition for video games and film. While, in the first case, the composition may be interactive because of the unpredictability of the player's actions while guaranteeing clarity of informations in the image, in the second case, the composition is designed upstream and is frozen on film, never to be changed upon viewing.

In the case of visual composition for real time, the question is raised of the use of expressive tools which can enter into conflict with the unpredictability of a performer's actions.

The case of video games is also already very clear concerning this issue, but it is even more so in the case of augmented scenes, in which new factors come into play. In our case study, the avatar is piloted by an actor on stage via motion capture, and the actors can interact with one another and create unforeseen situations. A digital artist or the director may decide to draw attention to details of the virtual stage, and the image projected onscreen is not necessarily designed ergonomically as in the case of video games and can extend itself toward an exploitation of cinematic tools that is more complete, but also has all the complexity of the unpredictable evolution of the avatar. The visual composition of the augmented scene is thus based on a set of unpredictable actions involving multiple human actors. Composing an augmented scene requires the upstream provision, therefore, of ways to modify the composition.

Before defining a free-interaction space for visual composition, we can define two forms of composition that we have used: fixed composition, and manipulated composition. The difference between these two types of composition is the extent of intervention by a person through tools enabling the transformation of the visual composition onscreen. The idea of every person on the stage having the ability to participate in composition enables the delegation of tasks, as well as the creation of expressive situations related to their roles. For example, it has been planned that the gestures of the mocaptor have an effect on elements of the scene, and thus potentially on the camera or lighting. The manipulactor takes charge of the avatar's position, and can also be responsible for the manipulation of other digital elements.

In this sense, in a fixed composition, the camera, lights, or elements of décor remain in a fixed position so as to guarantee readability of the image in the context of an onscreen projection. This composition relies above all on cinematographic camera work, and only the avatars can evolve freely. However, the parameters of a manipulated composition evolve with the actions of the different people in the scene. Thus, the camera, lights, or onscreen elements are in motion, and involve new visual composition as each event takes place. The table below (Figure 13) sums up the strengths and weaknesses of compositions currently used:

|  | Strengths | Weaknesses |
| --- | --- | --- |
| **Fixed composition** | + Use of a well-established photographic/cinematographic language<br>+ Control of plan | - Little freedom of improvisation<br>- Loss of interactive power |
| **Manipulated composition** | + Great capacity for improvisation and expression<br>+ Digital artist can perform rather than being a simple technician | - Precision is difficult<br>- Significant camera presence (temporality of camera placement) |

*Figure 12. Table showing strengths and weaknesses of the two types of composition used for augmented scenes in our system*

In our study of visual composition for staging a scene, we have attempted a fixed composition in order to exploit cinematic tools as far as possible (see Figure 14). While the camera and lights are placed in specific spots, the composition is then fixed and any modification will result in a new form of composition. The explanation we are seeking to give through this example allows us to emphasize the expressive responsibility of the tools we are developing for augmented scenes.

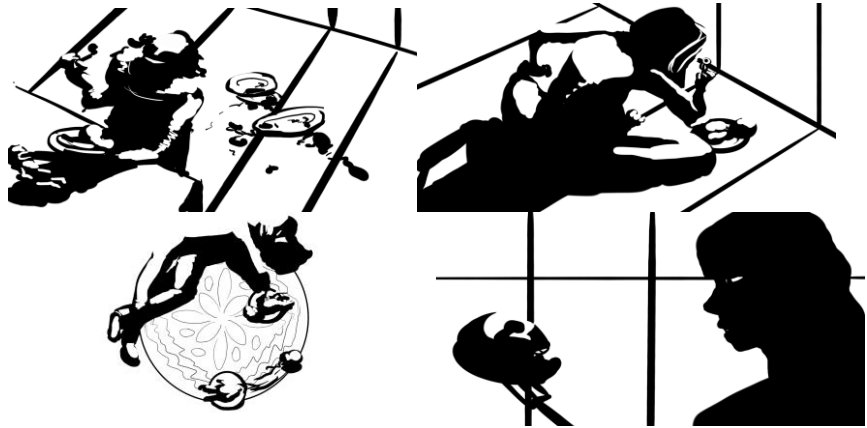

*Figure 13. Studies of visual compositions for cannibalism*

Our initial method will consist of hybridizing the traveling and storytelling techniques of film with the camera-movement tools used in video game development. A fixed composition would strip any interest in interacting with the image, but manipulated composition could lead to imprecision on the part of the person executing the action. Hypothetically speaking, to delineate the field of intervention in a visual composition, one possibility would be to use rails and intelligent camera observations.

### *1.5.3. Redefining the role of the digital artist*

It is necessary to specify a few things with regard to the role of the digital artist in the experimentation process. While this individual is not a physical actor or mocaptor, he is nevertheless involved in all the digital parameters of the manipulated visual composition, which he can then make available to the manipulactor: screen display, virtual cameras, virtual objects, avatar movement constraints, etc. He has an important power of action which he must continuously put in dialogue with the actors and director; he can, for example, make it possible to change the virtual camerawork and lighting to accentuate the avatar's expressions. He can also provide new movement potentialities to the neutral avatar via the mocaptor's body language.

As we have seen, there was a lack of common vocabulary available for use in terminological discussions of actor augmentation. This shortcoming is related to the crisis of indexicality invoked by Auslander concerning the relationship between the actor producing digital mocapture data and the avatar onto which this data is transferred [AUS 2017]. In the context of 3D film animation, Auslander indicates that the transformation of motion capture data by the digital artist in order to produce an avatar makes it difficult to assess the quality of the interpretation by the physical actor producing digital data.

In a theatrical context, we recognize that an actor's performance has to do not only with his quality of interpretation alone, but also with the instructions given by the director. The director's influence is not problematic; on the contrary. We believe, therefore, that the digital artist must be placed at the heart of the direction of the avatar, and a new avatar direction paradigm must be constructed that integrates new augmentation writing tools into a close and equal collaboration with three protagonists: the mocaptor, the manipulactor, the director, and the digital artists.

## 1.6. Conclusion

This article has described the impact of the use of a new portable motion-capture device in an ongoing theatrical experiment in which physical actors and digital avatars interact onstage. The new mocap system resembles a sort of tagging of the human body, which makes it possible to transcribe in real time a hybridization between physical and digital space via the intervention of an avatar manipulated by a mocaptor. We are proposing a method that enables the digital artist to articulate the data produced by the mocaptor in the digital avatar's frame of reference with the aim of producing a creative augmentation of gestures improvised live. This method relies on a highly adjustable software architecture, the characteristics of which we have described that are concerned by the challenges of motion capture and information exchange.

The real-time production of the avatar requires a close dialogue between the digital artist, the manipulactor, the mocaptor, and the director, which involves utilizing various types of feedback on the digital elements produced; we have described those already in use. The simplicity of use of the new mocap system has greatly stimulated the mocaptor and the director, which has led to requests for avatar augmentation requiring extensive expertise on the part of the digital artist. The result of this is that new challenges related to avatar direction must now be considered in an overlapping distribution of roles among all the protagonists involved.

Our next experimental prospectives will lead us toward the production and documentation of communication interfaces for this distribution of roles, which will enable the creation of a lexicon and vocabulary for the interactions in play.